\documentclass[aps,prX,superscriptaddress,twocolumn]{revtex4}
\usepackage{amsmath}
\usepackage{amsfonts}
\usepackage{graphicx}
\usepackage{color}
\usepackage{dsfont}
\usepackage{verbatim}
\usepackage{mathtools}
\usepackage[normalem]{ulem}
\usepackage{comment}

\definecolor{myblue}{rgb}{.93, .93, 1}

\newcommand{\bsub}{\begin{subequations}}
	\newcommand{\esub}{\end{subequations}}

\begin{document}
	
	\title{Superconductor versus insulator in twisted bilayer graphene}
	
	\author{Yang-Zhi~Chou}\email{YangZhi.Chou@colorado.edu}
		\affiliation{Department of Physics and Center for Theory of Quantum
		Matter, University of Colorado Boulder, Boulder, Colorado 80309,
		USA} 
	\author{Yu-Ping~Lin}
	\affiliation{Department of Physics and Center for Theory of Quantum
		Matter, University of Colorado Boulder, Boulder, Colorado 80309,
		USA} 
		\author{Sankar Das Sarma}
		\affiliation{Condensed Matter Theory Center and Joint Quantum Institute, Department of Physics, University of Maryland, College Park, Maryland 20742-4111, USA}
	\author{Rahul~M.~Nandkishore}
	\affiliation{Department of Physics and Center for Theory of Quantum
		Matter, University of Colorado Boulder, Boulder, Colorado 80309,
		USA} 
		
	\date{\today}
	
\begin{abstract}
We present a simple model that we believe captures the key aspects of the competition between superconducting and insulating states in twisted bilayer graphene. Within this model, the superconducting phase is primary, and arises at generic fillings, but is interrupted by the insulator at commensurate fillings. Importantly, the insulator forms because of electron-electron interactions, but the model is agnostic as to the superconducting pairing mechanism, which need not originate with electron-electron interactions. The model is composed of a collection of crossed one-dimensional quantum wires whose intersections form a superlattice. 
At each superlattice point, we place a locally superconducting puddle which can exchange Cooper pairs with the quantum wires.
We analyze this model assuming \textit{weak} wire-puddle and wire-wire couplings. We show that for a range of repulsive intrawire interactions, the system is superconducting at `generic' incommensurate fillings, with the superconductivity being `interrupted' by an insulating phase at commensurate fillings. 
We further show that the gapped insulating states at commensurate fillings give way to gapless states upon application of external Zeeman fields. These features are consistent with experimental observations in magic-angle twisted bilayer graphenes despite the distinct microscopic details. 
We further study the full phase diagram of this model and discover that it contains several {\it distinct} correlated insulating states, which we characterize herein.
\end{abstract}

\maketitle

\section{Introduction}

The discovery of correlated insulating phases and superconductivity in magic angle twisted bilayer graphene (TBLG) \cite{Cao2018tbg1,Cao2018tbg2,Yankowitz2019,Kerelsky2018,Cao2019strangeM,Lu2019Efetov,Jiang2019} and related heterostructures has triggered an explosion of interest in this phenomenon.  An ocean of ink has been spilled trying to explain this phase diagram - see e.g. \cite{po18prx,xu18prl, model5, gonzalez19prl,xie18ax, sherkunov18prb, isobe18prx, kennes18prb, LN, Kang2018prx,Zou2018prb,Yuan2018prb,Koshino2018prx,Wu2018,Lian2018twisted,Wu2019_phonon_linearT,Fu2018MAS,Lin2018KL,Classen2019,Peltonen2018,Wu2019Identification,Wolf2019electrically,Liu2019} for a representative but not exhaustive list. Most of these works have adopted the premise that the insulating phase, arising from electron correlation effects in the flatband TBLG, is `primary' and the superconducting phase descends from the insulator (a natural assumption given the original experimental data \cite{Cao2018tbg1, Cao2018tbg2}, where superconductivity arose only at particular fillings close to an insulating phase). More recent experimental data \cite{Lu2019Efetov}, however, has revealed superconductivity over a much broader range of dopings (with many more insulating states at commensurate fillings), and suggests a different perspective, where the superconductivity is `primary' and the system is superconducting at generic fillings, but where the superconductivity is `interrupted' by an insulating phase at commensurate fillings.  

In this work we introduce and study a simple model realizing superconductivity at generic fillings, interrupted by insulating states at commensurate fillings. Crucially, while the insulating states arise due to electron-electron interactions, the model is agnostic as to the pairing mechanism, which could be e.g. electron-phonon in origin \cite{Wu2018,Lian2018twisted,Peltonen2018,Wu2019Identification}. The model is highly simplified and does not pretend to be a detailed microscopic description of TBLG. However, we believe that it captures the key physics of the competition between the superconducting and insulating states in this system. Moreover the model is simple, well motivated, and exhibits a rich phase diagram, including several sharply distinct insulating phases, and as such should provide an interesting point of departure for future theoretical and experimental investigations of the observed TBLG phase diagram.

The model consists of a network of locally superconducting puddles connected by quantum wires. Such a model does not depend on the pairing mechanism at play - the basic assumption is simply that {\it local} superconducting fluctuations generically arise, for whatever reason, and we now wish to determine if the {\it global} phase structure is that of a superconductor (generic fillings) or an insulator (commensurate fillings). This depends on whether the quantum wires are able to efficiently mediate Josephson couplings between the puddles, which in turn depends on whether the wires are themselves in a metallic phase, and on whether Cooper pair tunneling between puddles and wires is a relevant perturbation. We address these questions using a Luttinger liquid approach, whereby intrawire interactions are treated non-perturbatively, but interwire and wire-puddle couplings are treated in a perturbative fashion. The system is assumed clean, although as we will subsequently argue, the results are expected to be robust to weak disorder. The resulting system is shown to have a rich phase diagram, with superconductivity at generic fillings, and multiple potential interaction-driven insulating phases at commensurate fillings. The magnetic field response of these insulating phases is discussed, and it is demonstrated that some at least are suppressed by Zeeman field, consistent with experimental observations \cite{Cao2018tbg1,Cao2018tbg2}.
The tantalizing similarity between our theoretically obtained `phase diagram' and the TBLG experimental observations should be taken seriously, and we believe that, in spite of the manifest simplicity of our model, our work may have captured some essential minimal aspects of the physics underlying the recent TBLG experimental observations.

A `coupled wire network' (see \cite{Meng2019}, and the references therein) of this sort was first proposed as an effective theory for twisted bilayer graphene in \cite{Efimkin2018}. The logic was that regions with locally AB registry would be gapped by an external electric field, but gapless modes would arise at the domain walls between locally AB and locally BA regions. These domain walls would intersect at regions of locally AA registry, which would also be gapless. A detailed microscopic model of this sort gives rise to a {\it triangular} lattice of quantum wires supporting {\it helical} modes. In this work, however, we simplify further to a {\it square} lattice of quantum wires hosting `ordinary' (not helical) Luttinger liquids, with no valley degeneracy. We believe this simplified model captures the key physics of the competition between superconductivity and insulating behavior, and as we will argue, we do not believe the differences in microscopic details to be important for this key physics. We note also that while the competition between superconductivity and insulating states in wire models has previously been discussed in \cite{Wu2019coupled-wire}, in this prior work the superconducting and insulating states both arise from electron-electron interactions within or between the quantum wires. In contrast, in our work the superconducting state can have a completely separate origin, and need not arise from the same mechanism as the insulating state with which it competes. Technically, this distinction is implemented by our placing locally superconducting puddles at the wire intersections. 
Since it is difficult to envision generic superconductivity arising from electron-electron interactions because of its fundamental repulsive nature, the flexibility of our model allowing superconducting and insulating states to arise from distinct mechanisms, if necessary, is a key feature of our theory.

The rest of this paper is organized as follows: We first introduce the interacting network model. Then, standard bosonization and weak coupling analysis are applied. We discuss all the possible phases and construct phase diagrams. We then discuss the implications for TBLG and potential generalizations, and conclude. 

\begin{figure}[t!]
	\includegraphics[width=0.35\textwidth]{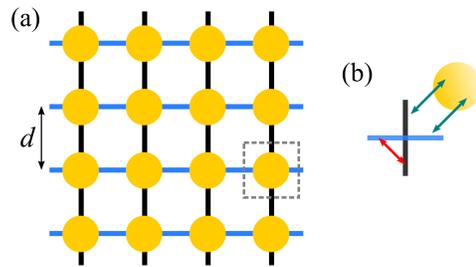}
	\caption{The setup of the hybrid network system. Black and blue lines indicate horizontal and vertical 1D quantum wires; yellow dots are the isolated superconducting puddles. (a) The network consists of horizontal and vertical quantum wires that intersect at the positions of the superconducting puddles. (b) Zoom-in at an intersection.  All the interwire (red arrows) and puddle-wire couplings (green arrows) are treated perturbatively. We perform Luttinger liquid analysis in this regime.}
	\label{Fig:SC_LL}
\end{figure}

\section{Model}

We consider a network consisting of zero-dimensional superconducting puddles and one-dimensional (1D) crossed quantum wires. As illustrated in Fig.~\ref{Fig:SC_LL}, the quantum wire arrays form a square superlattice with a lattice constant $d$, and the superconducting puddles are located at the superlattice points (wire intersections). In the absence of any coupling, the phases of the superconducting puddles are random. Meanwhile, the couplings at the wire intersections enable interaction-driven phases including two-dimensional (2D) superconductivity. This network model can be analyzed via standard bosonization analysis for 1D fermions \cite{Shankar_Book,Giamarchi_Book} in the weak wire-puddle and wire-wire couplings limit.

The network model in Fig.~\ref{Fig:SC_LL} can be decomposed to three parts: 1D quantum wires ($H^{(\text{1D})}$), superconducting puddles ($H^{(\text{SC})}$), and couplings at the wire intersections ($H^{(\text{X})}$). First, the quantum wire Hamiltonian is $H^{(\text{1D})}=\sum_{n_x}H^{(\text{1D})}_{y;n_x}+\sum_{n_y}H^{(\text{1D})}_{x;n_y}$ where $H^{(\text{1D})}_{y;n_x}$ ($H^{(\text{1D})}_{x;n_y}$) describes a vertical (horizontal) quantum wire located at $x=n_x d$ ($y=n_yd$) and $n_x$ ($n_y$) is an integer. 
We consider a single-channel spin$-\frac{1}{2}$ fermion for each 1D quantum wire. For simplicity, we assume that all the 1D systems are identical and with periodic boundary conditions. The predictions of this work will remain valid even away from this special limit. The kinetic part of the Hamiltonian is given by
\begin{align}\label{Eq:H_1D:H_0}
H_0=\sum_{\sigma=\uparrow,\downarrow}\int\limits_xv_F\left[R^{\dagger}_{\sigma}\left(-i\partial_xR_{\sigma}\right)-L^{\dagger}_{\sigma}\left(-i\partial_xL_{\sigma}\right)\right],
\end{align}
where $v_F$ is the Fermi velocity, $\sigma$ is the spin index, and $R_{\sigma}$ ($L_{\sigma}$) is the right (left) moving chiral fermion.

The interactions of interest are enumerated based on symmetry and relevance in renormalization group (RG) analysis. These are forward-scattering Luttinger liquid interactions and backscattering interactions. The former can be included nonperturbatively in the standard bosonization treatment \cite{Giamarchi_Book,Shankar_Book}.
Here, we only list the leading relevant backscattering interactions as follows \cite{Giamarchi_Book}:
\begin{subequations}\label{Eq:H_1D:int}
\begin{align}
\label{Eq:H_U}H_{U}=&U\int\limits_x:\left[e^{i\left(4k_{F}-Q\right)x}L^{\dagger}_{\uparrow}R_{\uparrow}L^{\dagger}_{\downarrow}R_{\downarrow}+\text{H.c.}\right]:,\\
\label{Eq:H_V}H_{V}=&V\int\limits_x:\left[L^{\dagger}_{\uparrow}R_{\uparrow}R^{\dagger}_{\downarrow}L_{\downarrow}+\text{H.c.}\right]:,
\end{align}
\end{subequations}
where $k_F$ is the Fermi wavevector, $Q$ is the wavevector commensurate to the superlattice, and $:\mathcal{O}:$ denotes the normal ordering of $\mathcal{O}$. $H_U$ corresponds to a umklapp interaction of the charge mode, and $H_V$ is a conventional backscatering of the spin mode. In this work, we assume the spin sector is gapped by this backscattering in the absence of an external magnetic field.

In addition to the intrawire interactions in quantum wires, one can apply an external Zeeman field to spin-polarize the system. The primary effect of the external magnetic field can be captured by the fermi wavevector mismatch between different spin species, $|k_{F\uparrow}-k_{F\downarrow}|=\delta k_F\propto B$ ($B$ the strength of the Zeeman field). This generally suppresses the spin gap; the charge commensuration is less sensitive to the Zeeman field unless the 1D band curvature becomes important.

Now, we discuss the zero-dimensional superconducting puddle. $H^{(\text{SC})}=\sum_{n_x,n_y}H^{(\text{SC})}_{n_x,n_y}$, where $H^{(\text{SC})}_{n_x,n_y}$ is a local superconducting puddle at $(x,y)=(n_xd,n_yd)$.
For each $n_x$ and $n_y$,
\begin{align}\label{Eq:H_SC}
H^{\text{(SC)}}_{n_x,n_y}=\sum_{j}E_j\left(d^{\dagger}_{j\uparrow}d_{j\uparrow}+d^{\dagger}_{j\downarrow}d_{j\downarrow}\right)-G\sum_{j,j'}d^{\dagger}_{j\uparrow}d^{\dagger}_{j\downarrow}d_{j'\downarrow}d_{j'\uparrow},
\end{align} 
where $E_j$ is the energy of level $j$ and $G>0$ is the interaction coupling constant. We consider local mean-field s-wave superconducting order of the puddle which
requires two assumptions \cite{Spivak2008,Nandkishore2013}: (a) The local temperature is lower than the critical temperature, and (b) the size of the puddle is larger than the coherence length. We further assume that the local order parameter is $\Delta_{n_x,n_y}=-G\sum_j\langle d_{j\uparrow}d_{j\downarrow} \rangle_{n_x,n_y}=|\Delta|e^{i\Phi_{n_x,n_y}}$, where $|\Delta|$ is a constant amplitude and $\Phi_{n_x,n_y}$ is the position-dependent phase. A uniform superconductivity is established if the system develops \textit{global} phase rigidity, but not if the phase on each puddle can be freely varied. 

Last, we discuss all possible perturbations (allowed by symmetry and based on relevance in RG analysis) at the wire intersections (superlattice points). 
There are two types of couplings: wire-puddle couplings $H^{(\text{X})}_{\text{wp}}$ and wire-wire couplings $H^{(\text{X})}_{\text{ww}}$. Since each puddle forms a superconducting gap, the primary wire-puddle coupling is the Cooper pair hopping. Such a process at each $n_x$ and $n_y$ is given by
\begin{align}\label{Eq:H_X:wp}
H^{\text{(X)}}_{\text{wp}}=\left[J_xd^{\dagger}_{0\downarrow}d^{\dagger}_{0\uparrow}\mathcal{C}_{x}+J_yd^{\dagger}_{0\downarrow}d^{\dagger}_{0\uparrow}\mathcal{C}_{y}+\text{H.c.}
\right],
\end{align}
where $J_x$ and $J_y$ are the Cooper pair hopping strengths, $d_{0\sigma}$ is the puddle electron with spin $\sigma$ at the Fermi energy, and $\mathcal{C}_x$ ($\mathcal{C}_y$) is the spin-singlet Cooper pair operator of the horizontal (vertical) quantum wire. The explicit expression for the Cooper pair hopping is given by $\mathcal{C}_{x,y}=\left[L_{\downarrow}R_{\uparrow}-L_{\uparrow}R_{\downarrow}\right]_{x,y}$. Other wire-puddle interactions are irrelevant and thereby ignored in this work.

There are three primary processes that couple horizontal and vertical quantum wires at the wire intersections: single electron tunnelings ($H_e$), charge-density-wave interactions ($H_C$), and Cooper pair hoppings ($H_{2e}$). For each $(n_x,n_y)$, the wire-wire coupling Hamiltonian is $H^{\text{(X)}}_{\text{ww}}=H_{e}+H_{\text{C}}+H_{2e}$, where
\begin{subequations}\label{Eq:H_X:ww}
\begin{align}
H_{e}=&t_e\sum_{\sigma}\left[\left(R^{\dagger}_{\sigma}+L^{\dagger}_{\sigma}\right)_y\left(R_{\sigma}+L_{\sigma}\right)_x+\text{H.c.}\right],\\
H_{C}=&t_u\sum_{\sigma,\sigma'}\!:\!\left(L^{\dagger}_{\sigma}R_{\sigma}+R^{\dagger}_{\sigma}L_{\sigma}\right)_y\!\left(L^{\dagger}_{\sigma'}R_{\sigma'}+R^{\dagger}_{\sigma'}L_{\sigma'}\right)_x\!:,\\
H_{2e}=&t_{2e}\left[\left(R^{\dagger}_{\uparrow}L^{\dagger}_{\downarrow}-R_{\downarrow}^{\dagger}L^{\dagger}_{\uparrow}\right)_y\left(L_{\downarrow}R_{\uparrow}-L_{\uparrow}R_{\downarrow}\right)_x+\text{H.c.}\right].
\end{align}
\end{subequations}
The phases of the network model depend on the above junction perturbations. We turn to analyze the perturbations and construct the phase diagram next.

\section{Analysis}

The network model can be studied analytically in the weak wire-puddle and wire-wire coupling limit, with intra-wire interactions being treated non-perturbatively. 
The analysis in this section is strictly valid when $L\gg d \gg \alpha$, where $L$ is the system size, $d$ is the superlattice constant, and $\alpha$ is the microscopic scale, which in turn requires the twist angle to be small. 
We first review the properties of the quantum wires in the standard bosonization approach. Then, the wire-puddle and wire-wire couplings are treated as perturbations. 
We further assume that the amplitude of the local superconducting order parameter in a puddle cannot be suppressed but the phase can be modified. Using these approximations, we can classify the phases in our interacting network model. Our objective here is to capture the effects of commensuration and external Zeeman fields. 
The main results are summarized in Table~\ref{Table:phases} and Fig.~\ref{Fig:PD}.

\subsection{Quantum wires as Luttinger liquids}\label{Sec:1Dwire}

Each quantum wire can be described by a spinful Luttinger liquid. With the standard bosonziation \cite{Shankar_Book,Giamarchi_Book}, the charge and the spin collective modes are decoupled. The 1D Hamiltonian $H_0+H_U+H_V$ [given by Eqs.~(\ref{Eq:H_1D:H_0}) and (\ref{Eq:H_1D:int})] becomes $H_{c}+H_{s}$, where
\begin{align}
\nonumber H_{c}=&\frac{v_c}{2\pi}\int\limits_x\left[K_c\left(\partial_x\phi_c\right)^2+\frac{1}{K_c}\left(\partial_x\theta_c\right)^2\right]\\
\label{Eq:H_c}&+\frac{U}{2\pi^2\alpha^2}\int\limits_x\cos\left[2\sqrt{2}\theta_c+\left(4k_{F}-Q\right)x\right],\\
\nonumber H_{s}=&\frac{v_s}{2\pi}\int\limits_x\left[K_s\left(\partial_x\phi_s\right)^2+\frac{1}{K_s}\left(\partial_x\theta_s\right)^2\right]\\
\label{Eq:H_s}&+\frac{V}{2\pi^2\alpha^2}\int\limits_x\cos\left[2\sqrt{2}\theta_s\right].
\end{align}
In the above expressions, $\theta_c$ ($\theta_s$) is the `phonon-like' boson field of the charge (spin) collective mode, $\phi_c$ ($\phi_s$) is the `phase' field of the charge (spin) collective mode, $v_c$ ($v_s$) is the velocity of the charge (spin) mode, $K_c$ ($K_s$) is the charge (spin) Luttinger liquid parameter, and $\alpha$ is the microscopic ultraviolet length scale. The bosonization convention is summarized in Appendix~\ref{App:Bosonization}.
In the absence of Luttinger liquid interactions, $v_c=v_s=v_F$ and $K_c=K_s=1$. $K_c<1$ ($K_c>1$) for repulsive (attractive) charge interactions. We consider general cases where $K_c$ and $K_s$ can be tuned arbitrarily. The results for specific microscopic models can be extracted from our analysis.
Note that we are assuming short-range electron-electron interactions in our bosonization calculations, which is a standard approximation for generic Luttinger liquid theories.  This can be justified by assuming that some screening of Coulomb interaction necessarily happens in TBLG because of the presence of metallic gates in the system.  Actually, our theory should apply for long-range Coulomb interactions also with the modification that the Luttinger exponents become effective scale-dependent exponents falling off extremely slowly, slower than any power laws, at long distances.

The charge sector will remain gapless as long as $|4k_{F}-Q|>\delta Q_c$ \cite{PokrovskyTalapov} ($\delta Q_c$ is the critical threshold). In this puddle-wire network model, the commensuration wavevector $Q=0,\pm 2\pi/d$ where $d$ is the superlattice constant.
With commensuration ($|4k_{F}-Q|<\delta Q_c$), the perturbative RG flows \cite{Giamarchi_Book} are
\begin{align}
\label{Eq:SG_RG_U}\frac{dy_U}{dl}=&\left(2-2K_c\right)y_U,\\
\label{Eq:SG_RG_K}\frac{d K_c}{dl}=&-\frac{1}{2}y_U^2K_c^2,
\end{align}
where $y_U=U/(\pi v_c)$.
For $K_c<1$, $H_U$ at commensuration is relevant. 
For $K_c>1$, $H_U$ is irrelevant for infinitesimal $|U|$. Since the Eqs.~(\ref{Eq:SG_RG_U}) and (\ref{Eq:SG_RG_K}) form hyperbolic trajectories,
$|y_U|$ can still flow up (implying a formation of a charge gap) when $|y_U|$ is above the threshold value (separatrix).

Meanwhile, $H_V$ given by the cosine term in Eq.~(\ref{Eq:H_s}) is a regular backscattering contribution. 
With $K_s<1$ or a sufficiently large $|V|$, the spin sector $H_{s}$ becomes fully gapped at zero temperature. This is the same RG flow as that in the charge sector with commensuration. 
In addition, the phase field of spin, $\phi_s$, will be completely disordered. In this work, we will largely focus on the network built by spin-gapped quantum wires.

We thus see that commensuration \textit{naturally} has the effect of gapping out the quantum wires connecting superconducting puddles. When said wires are gapped, they cannot mediate a long range Josephson coupling between puddles (the Josephson coupling will decay exponentially with distance \cite{Nandkishore2013}). As a result, the system cannot be a global superconductor (at least at experimental temperatures), and moreover, being everywhere gapped, will exhibit insulating behavior. This gives a natural mechanism for insulating phases to emerge at commensuration - [Fig. 2(b)]. 

We now discuss the effect of an external magnetic field (Zeeman field). For simplicity, we consider a field applied {\it in plane} which will not couple to the orbital degree of freedom, but will enter only through the Zeeman coupling. As discussed in the previous section, the Fermi points of different spin species shift in the opposite way. The cosine term in Eq.~(\ref{Eq:H_s}) becomes $\cos\left[2\sqrt{2}\theta_s+2\left(k_{F\uparrow}-k_{F\downarrow}\right)x\right]$. The spin sector will now remain gapless for a sufficiently large $|k_{F\uparrow}-k_{F\downarrow}|$. With linear dispersion, the shifts of the Fermi points do not affect $k_{F\uparrow}+k_{F\downarrow}$, so the charge sector remains the same as in the absence of a magnetic field. With a finite curvature in the 1D band, the charge sector at the commensurate filling can also become gapless. To understand the implications for the global phase, however, we must also analyze the effect of interwire and wire-puddle couplings. 

\subsection{Couplings at intersections}\label{Sec:X_RG}

Here, we discuss the roles of wire-puddle and interwire perturbations at the wire intersections given by Eqs.~(\ref{Eq:H_X:wp}) and (\ref{Eq:H_X:ww}). To simplify the perturbative analysis, we consider \textit{static} superconducting puddles. 
Under such an approximation, we can derive perturbative RG flows for all the perturbations at the intersections. 

The wire-puddle coupling is dictated by hopping of Cooper pairs. Under the static assumption, we can replace $d^{\dagger}_{0\downarrow}d^{\dagger}_{0\uparrow}$ at $(n_x,n_y)$ by its expectation value $\langle d_{0\downarrow}^{\dagger}d_{0\uparrow}^{\dagger} \rangle_{n_x,n_y}$ in Eq.~(\ref{Eq:H_X:wp}). The spin-singlet Cooper pair operator ($\mathcal{C}$) in a quantum wire is bosonized to
\begin{align}
\mathcal{C}
=&\frac{1}{\pi\alpha}\hat{U}_{\downarrow}\hat{U}_{\uparrow}e^{i\sqrt{2}\phi_c}\cos\left(\sqrt{2}\theta_s
\right),
\end{align}
where $\hat{U}_{\sigma}$ is the Klein factor of an electron carrying spin $\sigma$.
With perturbative analysis in bosonization, the wire-puddle Cooper pair hopping gives an RG flow as follows
\begin{align}\label{Eq:RG:J_xy}
\frac{d J_{x,y}}{dl}=&\left(1-\frac{1}{2K_c}-\frac{K_s}{2}\right)J_{x,y}.
\end{align}
The wire-puddle coupling $J_{x,y}$ becomes relevant when $1/K_c+K_s<2$. In order for the system to develop global superconductivity, we need $J_{x,y}$ to be relevant. For spin-gapped quantum wires, we take $K_s\rightarrow 0$ and treat $\cos\left(\sqrt{2}\theta_s\right)$ as a constant.
The corresponding RG flow suggests that Cooper pair hopping becomes relevant when $K_c>1/2$. 
This means that the pairing at the wire intersection survives even for repulsively interacting spin-gapped quantum wires ($1/2<K_c<1$). Meanwhile, the Cooper pair hopping is suppressed parametrically for $V>0$ because the expectation value of $\cos\left(\sqrt{2}\theta_s\right)$ is small \cite{fn1}. 
We note that $J_{x,y}$ is always irrelevant for charge-gapped quantum wire (equivalently to $K_c\rightarrow 0$) as the $\phi_c$ term becomes completely disordered. 

The interwire couplings [given by Eq.~(\ref{Eq:H_X:ww})] can also be studied via standard perturbative RG methods and bosonization. The bosonized expressions are summarized in Appendix~\ref{App:RG_flow:imp}.
We write down the leading order RG flows as follows:
\begin{align}
\frac{dt_e}{dl}=&\left[1-\frac{1}{4}\left(K_c+\frac{1}{K_c}\right)-\frac{1}{4}\left(K_s+\frac{1}{K_s}\right)\right]t_e,\\
\frac{dt_u}{dl}=&\left(1-K_c-K_s\right)t_u,\\
\frac{dt_{2e}}{dl}=&\left(1-\frac{1}{K_c}-K_s\right)t_{2e}.
\end{align}
$t_e$ is the single electron tunneling and is at most marginal when $K_c=K_s=1$. In the presence of Luttinger interactions ($K_c\neq 1$ and/or $K_s\neq 1$), we can just ignore single electron tunnelings for small $|t_e|$ since this operator is irrelevant. Meanwhile $t_u$ is the density-density backscattering at the intersection which is relevant for $K_c+K_s<1$. When $t_u$ becomes relevant,
the systems develop `clogged junctions' \cite{Chou2019}, inducing a globally insulating phase. Finally, $t_{2e}$ describes the Cooper pair hoppings among two intersecting quantum wires. Such a process becomes relevant when $1/K_c+K_s<1$. We note that both $K_c$ and $K_s$ are unaffected by the perturbations at the intersections since we have taken the limit $d\gg \alpha$.

For the spin-gapped quantum wires, the RG flows can be obtained by setting $K_s\rightarrow 0$. In this case, $t_e$ and $t_{2e}$ are always irrelevant due to  completely disordered $\phi_s$. Based on the perturbative RG, $t_u$ becomes relevant when $K_c<1$. However,
the $t_u$ interaction with $V>0$ is parametrically suppressed \cite{fn1}, same as the suppression in the wire-puddle Cooper pair hopping. Similar analysis for charge-gapped quantum wires can be done by setting $K_c\rightarrow 0$.

The properties of these wire-puddle and interwire interactions dominate the network phases. In particular, the effective Josephson coupling determines if the system can form uniform 2D superconductivity. There are other interaction-driven phases from this network model. We will turn to construct the phase diagrams next.

\subsection{Phase diagram}\label{Sec:PD}

\begin{table*}[]
	\begin{tabular}{c||c|c||c|c|c}
		Phase & 1D C-gap & 1D S-gap & X pairing & X C-gap & X S-gap\\ \hline\hline
		SC/SC$^*$ & $\mathsf{x}$  & - & $\checkmark$ & $\mathsf{x}$ & - \\ \hline
		Fully Gapped & $\checkmark$ & $\checkmark$ & $\mathsf{x}$ & $\checkmark$ & $\checkmark$\\ \hline
		C-Clogged & $\mathsf{x}$ & $\checkmark$ &  $\mathsf{x}$ & $\checkmark$ & $\checkmark$\\ \hline
		S-Clogged & $\checkmark$ & $\mathsf{x}$ &  $\mathsf{x}$ & $\checkmark$ & $\checkmark$\\ \hline
		Metal (M) & $\mathsf{x}$ &$\mathsf{x}$ & $\mathsf{x}$ & $\mathsf{x}$ & $\mathsf{x}$\\ \hline
		S-LL  & $\checkmark$ & $\mathsf{x}$  & $\mathsf{x}$ & $\checkmark$ & $\mathsf{x}$
	\end{tabular}
	\caption{Possible network phases and the corresponding properties. 1D C-gap (S-gap) indicates the gap in the charge (spin) sector of 1D quantum wires; X pairing means a finite Cooper pairing tunneling at the intersection, characterized by a relevant $J_{x,y}$; X C-gap (S-gap) corresponds to a local charge (spin) gap at the intersection. SC/SC$^*$ corresponds to a 2D superconducting phase; Fully gapped indicates an insulator with all the gapped wires and the gapped intersections; C-Clogged (S-Clogged) corresponds to an insulator with gapless charge (spin) 1D wires but with gapped intersections; Metal (M) indicates a metallic state; S-LL is a phase with 1D spin Luttinger liquids. These phases are discussed in the main text.}
	\label{Table:phases}
\end{table*}

In this network model, the zero temperature phases can be built based on the properties of the 1D quantum wires and the wire intersections. In the absence of any coupling, the system is just a collection of decoupled quantum wires and isolated superconducting puddles. We assume that the superconducting puddles have random phases in this case, corresponding to no global superconductivity. 
We now discuss whether phase rigidity and global superconductivity should develop given the couplings at the wire intersections and the quantum wire properties. 

To realize a uniform superconductivity in the network model, the global phase coherence among the superconducting puddles is required. The charge conduction of the quantum wires is crucial as the superconducting puddles can interact by transferring Cooper pairs via the 1D quantum wires.
Therefore, a 2D superconducting state requires two conditions, gapless charge modes in the 1D quantum wires and relevant wire-puddle couplings ($J_{x,y}$) at wire intersections. 
Based on the perturbative RG analysis in the previous subsection, we can identify the regime that hosts uniform 2D superconductivity at zero temperature. Interestingly, the 2D superconductivity may survive repulsive interactions for spin-gapped quantum wires. There are five additional possible phases in this interacting network model, besides the 2D superconductor, and four of these are charge insulators - the fully gapped insulators, clogged insulators \cite{Chou2019},  and spin Luttinger liquid. The properties of the phases are summarized in Table~\ref{Table:phases}.

A fully gapped insulator requires both charge and spin gaps in the 1D quantum wire. This requires repulsive interactions ($K_c, K_s <1$) at weak backscattering, plus commensuration. Concomitantly, in this phase the Cooper pair hoppings at the wire intersections will be suppressed since $\phi_c$ becomes completely disordered. A clogged insulator in contrast contains gapless charge and/or spin modes in the quantum wires. At the wire intersections, both charge and spin modes are gapped by the `clogging' phenomenon discussed in \cite{Chou2019}. In Table~\ref{Table:phases}, we use C-Clogged and S-Clogged to distinguish the 1D gapless charge mode and spin mode respectively. These clogged phases can arise only for sufficiently strong intrawire repulsions, but can arise even away from commensuration. The clogged phases, like the fully gapped phase, are insulators for both charge and spin. Meanwhile, the spin Luttinger liquid phase is an insulator for charge, but a conductor for spin, and can only arise in the presence of commensuration {\it and} Zeeman field (or alternatively, with commensuration and interactions that are repulsive in the charge sector but \textit{weakly} attractive in the spin sector). Finally, in the presence of Zeeman field (or with \textit{weak} attractions that are attractive in the spin sector, $K_s >1$), there can arise a metallic phase which is a conductor for both charge and spin.

\begin{figure}[t!]
	\includegraphics[width=0.45\textwidth]{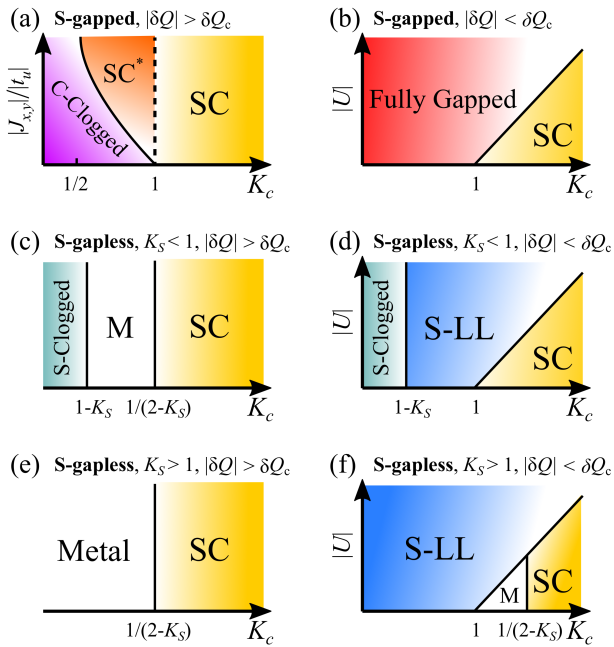}
	\caption{Zero temperature phase diagrams of the network model. The right column is near commensuration, the left column is away from commensuration. The properties of the phases are discussed and summarized in Table~\ref{Table:phases}. We present the phase diagrams with the 1D quantum wire properties. The spin gapped cases, (a) and (b), can be viewed as the absence of a Zeeman field. $\delta Q$ is the incommensuration measure in the charge umklapp interaction given by Eq.~(\ref{Eq:H_U}).
	For spin gapless cases, (c)-(f), we discuss $K_s<1$ [(c) and (d)] and $K_s>1$ [(e) and (f)] separately. We assume $0<K_s<2$ in this work.
	}
	\label{Fig:PD}
\end{figure}

We first focus on the network made by spin-gapped quantum wires. For a generic incommensurate filling [see Fig.~\ref{Fig:PD}(a)], the system is generically a superconductor, but can become a clogged insulator for sufficiently strong repulsive interactions. The superconducting state competes with the clogged state for $1/2<K_c<1$. 
We note that both the $t_u$ interaction and the wire-puddle Cooper pair tunneling are parametrically suppressed (but nonzero) for repulsive spin-gap interactions ($V>0$) \cite{fn1}. 
The clogged state dominates for $K_c<1/2$. With attractive interactions ($K_c>1$), the network model develops a 2D superconducting phase.
When the 1D spin gapped quantum wires are at commensurate electron fillings [see Fig.~\ref{Fig:PD}~(b)], the charge sector becomes gapped for $K_c<1$ or for sufficiently large $|U|$ \cite{Giamarchi_Book}. The fully gapped regime comes from the RG analysis in Eqs.~(\ref{Eq:SG_RG_U}) and (\ref{Eq:SG_RG_K}). As shown in Fig.~\ref{Fig:PD}(b), the superconducting regime is largely suppressed comparing with Fig.~\ref{Fig:PD}~(a). 
Combining Figs.~\ref{Fig:PD}(a) and \ref{Fig:PD}(b), we can identify a sizable range of states that experience superconductor-insulator transition tuned by the filling of the 1D quantum wires. This feature is reminiscent of the experimental observation of the superconductor-insulator transitions in \cite{Lu2019Efetov}. We will discuss the potential relations in the next section.

Now, we discuss the spin gapless quantum wires which can be viewed as the original spin gap wires in the presence of Zeeman fields. For a generic incommensurate filling [see Fig.~\ref{Fig:PD}(c) and \ref{Fig:PD}(e)], there are three possible phases: metallic phase, superconducting phase, and clogged insulators (S-Clogged). In this case, the 2D superconductivity takes place when $K_c>1/(2-K_s)$. For $K_s<1$, the clogged insulators appear for sufficiently strong charge repulsive interactions $K_c<1-K_s$.
At the commensurate fillings [see Fig.~\ref{Fig:PD}(d) and \ref{Fig:PD}(f)], a charge gap is formed in each spin-gapless quantum wire for $K_c<1$ or for sufficiently large $|U|$ \cite{Giamarchi_Book}. In the absence of other perturbations, the network phase is a collection of decoupled spin Luttinger liquids. A clogged insulator also arises for $K_c<1-K_s$, similar to the cases with incommensurate fillings. For $K_s>1$, there exists a region such that neither the charge gap nor the Cooper pair hopping is formed. Therefore, we predict a metallic phase between the spin Luttinger liquid phase and the 2D superconductivity. The clogged states do not arise for $K_S>1$.

\section{Correlated insulator phenomenology}\label{Sec:CI_Ph}

We first focus on the spin-gapped quantum wire network without commensurate fillings [Fig.~\ref{Fig:PD}(a)], which we believe to be the `generic' scenario. The 2D superconductivity appears generically for interacting quantum wires with attractive charge-charge interactions. More interestingly, the 2D superconductivity can even coexist with a repulsively interacting network for some parameter regions (SC$^*$ in Table.~\ref{Table:phases} and Fig.~\ref{Fig:PD}) \cite{fn3}. 
The establishment of the coherent superconducting phase is due to the virtual Cooper pair exchange among superconducting puddles via charge conducting quantum wires. Formally, one can derive an effective action of the coupling between superconducting puddles by integrating over the electronic degrees of freedom. In what follows, we summarize the main results and relegate the derivations to Appendix~\ref{App:EJC}.
The effective Josephson coupling depends on the property of the quantum wire network and is temperature dependent. 
The transition temperature ($T_c$) of this 2D superconductivity is determined by the nearest-neighbor effective Josephson couplings at $T_c$. For $T\ll T_c$, two distant superconducting puddles (separated by $r$) interact in a power law manner, $r^{\frac{1}{K_c}-1}$ for $K_c<1$ (repulsive interacting wires). For $K_c>1$ (attractive interacting wires), we have infinite range interactions such that the system should be well described by an infinite range mean field theory. In this regime, we do not expect finite range perturbations or spatial inhomogeneity (e.g., disorder) to have any significant impact.

For a spin gapped quantum wire network at commensurate fillings [Fig.~\ref{Fig:PD}(b)], fully gapped insulating states arise for repulsive interactions ($K_c<1$) or for a large $|U|$. The presence of charge gap blocks Cooper pair hopping between superconducting puddles. As a result, uniform 2D superconductivity is not expected owing to the incoherent phases among the puddles. 
The commensuration condition 
corresponds to filling factor $\nu=\pm 1/2$ ($\nu = \pm 2$ in the terminology of \cite{Lu2019Efetov}) in the moir\'e mini band. To see this, note that if the only gapless states are on the domain walls, then $2k_F$ for the domain walls must be proportional to density (Luttinger's theorem). When the conduction (valence) band is completely full (empty) then $2k_F$ must equal a reciprocal lattice vector. Ergo, when the conduction (valence) band is half full, then $2k_F$ must equal half a reciprocal lattice vector, such that the umklapp process is commensurate. Meanwhile, at charge neutrality $\nu=0$, we have $k_F = 0$ by particle hole symmetry, which also corresponds to a commensurate filling.  Such a filling-driven superconductor-insulator transition in the network model is reminiscent of that in the magic-angle twisted bilayer graphenes \cite{Cao2018tbg2,Yankowitz2019,Lu2019Efetov}. In particular, series of insulating states were observed at commensurate fillings, and the system stayed superconducting for generic fillings \cite{Lu2019Efetov}. Our theory provides a natural explanation for this observation.

Now, we discuss the effect of in-plane magnetic field (Zeeman field) on the spin-gapped quantum wire network.
Generically, the spin gap is sensitive to the applied Zeeman field as we have discussed in Sec.~\ref{Sec:1Dwire}.
The Zeeman field induces spin imbalance, $k_{F\uparrow}\neq k_{F\downarrow}$, such that the spin backscattering given by Eq.~(\ref{Eq:H_V}) becomes irrelevant \cite{fn2}. At commensurate fillings, the insulating regions in the phase diagram of the spin-gapped quantum wires (the absence of a Zeeman field) are thereby replaced by the spin Luttinger liquids or the clogged insulators [Fig.~\ref{Fig:PD} (d) and (f)]. 
We expect this effect to take place even if the superconducting puddles survive the applied magnetic field. A sufficiently strong Zeeman field may also suppress the charge gap due to finite 1D band curvature.
We note that a Zeeman field driven instability of the correlated insulator is also observed in TBLG experiments \cite{Cao2018tbg1,Cao2018tbg2,Yankowitz2019,Lu2019Efetov}. 
All the possible phases (ignoring suppression of superconducting puddles by Zeeman field) are summarized in Fig.~\ref{Fig:PD}(c)-(f). 

\section{Discussion}
In this work, we present a minimal network model that we believe captures a key aspect of the competition between superconductivity and insulating phases in TBLG. The model consists of quantum wires and superconducting puddles (see Fig.~\ref{Fig:SC_LL}). Within the standard weak coupling analysis, we demonstrate commensuration-driven superconductor-insulator transitions and Zeeman field suppression of the insulating states. Both features are reminiscent of the phenomenology in the magic-angle twisted bilayer graphene. We also construct the full phase diagram of this minimal model, which contains multiple competing correlated insulating phases. 

We now discuss how our minimal model relates to actual TBLG. 
As has been discussed \cite{Efimkin2018} and shown \cite{Yoo2019_reconstruction,Huang2018Helical_TBLGexp,Xu2019_GiantOscillation} in the literature, it is natural to expect the experimental systems with small twist angles to be almost everywhere gapped, with gapless states arising only at domain walls between regions with locally AB and locally BA registry, and at the vertices where such domain walls intersect, which will have locally AA registry. Systems at the first magic angle also share the above discussed properties except that the local AA registries contain considerable electronic states \cite{Yoo2019_reconstruction}.
Thus, a `coupled wire' network with `puddles' at the vertices should be a good description for TBLG. {\it However}, there are some differences in microscopic details. For starters, the wire network appropriate for TBLG would consist of a {\it triangular} lattice of quantum wires, with six wire segments connecting to each vertex, whereas we have simplified to a {\it square} lattice of quantum wires, with four wire segments connecting to every vertex. We do not believe that this detail will qualitatively affect the competition between superconducting and insulating phases. It will however complicate the analysis, and may affect some details regarding exactly where the phase transitions lie - we defer considerations of this more complicated lattice geometry to future work. Additionally, in physical TBLG, the quantum wires are {\it valley degenerate}, and contain {\it helical} domain walls, with the two valleys supporting domain walls with opposite helicity. In this work, we have ignored the valley degree of freedom altogether, and have modeled the quantum wires as ordinary (not helical) Luttinger liquids. We do not believe that ignoring the helical nature of the domain walls changes anything important since the `umklapp' interaction involving scattering two electrons from one valley to another conserves crystal momentum, modulo reciprocal lattice vectors of the moir\'e superlattice, and is thus allowed. The insulating phases in our analysis arise from the umklapp interaction. It is however possible, perhaps even likely, that taking valley degeneracy into account may lead to the emergence of still more potential insulating phases, including analogs of spin and valley polarized `quantum Hall ferromagnet' phases. Such physics is presumably crucial to the emergence of insulating phases at fillings $\nu = \pm1, \pm3$ in the terminology of \cite{Lu2019Efetov} (without taking valley physics into account, our analysis only predicts insulating phases as $\nu = 0, \pm2$). Exploration of the interplay of spin-valley degeneracy and polarization with the physics we have discussed herein would be a particularly interesting topic for future work. It would also be interesting to better understand the effect of the substrate, which has been ignored in the present analysis. 

Another essential simplification employed in this work is that we have assumed the locally superconducting puddles to involve {\it s-wave} pairing. If the local pairing symmetry was anything other than s-wave, then there would be scope for frustration in the Josephson couplings, and a more detailed analysis akin to \cite{Spivak2008,Moore2004} would be required. Conceivably, one could then obtain some kind of `pseudogap' phase whereby Josephson couplings between puddles were strong, but the system was prevented from establishing global phase coherence by frustration in the couplings. Since the nature of the pairing symmetry in TBLG is still not clear, we have steered clear of such an analysis at present, but it would be an interesting question to revisit in the future, once a consensus emerges as to the pairing symmetry in the system. 

In addition, we have approximated by treating the system as {\it clean}, whereas in reality the experimental system is known to be disordered. We do not expect weak disorder to qualitatively affect our results. The insulating phases, being gapped either in the wires or at the junctions, will be robust to disorder, whereas our superconducting phase is governed by long range Josephson couplings, and should thus also be robust as long as the disorder is weak. The effect of strong disorder is a more interesting problem, which could be studied in the spirit of Ref.~\cite{Lobos2012} (although, the TBLG samples of experimental interest being very clean with very high mobilties, the strong disorder limit may be of little practical relevance). Strong impurity backscattering will create a `bottleneck' for the charge transport in the associated wire segment.
A uniform 2D superconductivity can still arise as long as such bottleneck segments do not percolate the entire system. Another relevant form of disorder is the twisted angle fluctuation \cite{Beechem2014rotational,Yankowitz2019} which results in a modulation of the superlattice spacing $d$. Since the effective nearest-neighbor Josephson coupling depends on $d$, we expect phase separation of the superconducting domains and the normal state domains at some intermediate finite temperatures. Nevertheless, global superconductivity will persist as long as superconducting domains percolate.
Last, we have assumed that disorder is not strong enough for localization physics to become important. If it does, then that opens a whole new direction for exploration \cite{NS2017}. We defer a more detailed exploration of strong disorder physics to future work. 

Finally, we discuss the possible role of the `magic angle' twist in our theory. In fact, we believe the magic angle concept is \textit{not} essential to the experimentally observed TBLG phenomenology and all that is necessary is a twist angle small enough to substantially enhance interaction and correlation effects through the flatband moir\'e superlattice physics.  In our theory, the small twist angle leads to the creation of the superconducting puddles by strongly enhancing the applicable pairing interaction-- for electron-phonon interaction mediated TBLG superconductivity, for example, this is demonstrated explicitly in Refs. \cite{Wu2018,Lian2018twisted,Wu2019_phonon_linearT}. For larger twist angles, the superconducting puddles would not form and our model for competing superconducting and insulating phases would not apply.  Another aspect where the magic angle concept may apply is through the Luttinger parameters, which may all be close to unity (non-interacting limit) far away from the magic angle, whence the relevant cosine terms may be rendered only weakly relevant, thus becoming unimportant on experimental length scales. Our network model and the derived consequences, e.g. the competing quantum phases, while depending quantitatively on the precise value of the twist angle, are not crucially dependent on the twist angle being the exact magic angle. We simply need a twist angle small enough to enhance interaction effects sufficiently to lead to superconducting puddles and strongly relevant Luttinger parameters.

To conclude: we have provided a simple model which we believe captures the key physics of the competition between superconductivity and insulating behavior in twisted bilayer graphene. Crucially, the model is agnostic as to the origin of pairing, which need not come from electron-electron interactions, but nonetheless gives rise to a phase diagram with superconductivity at `generic' filings interrupted by correlation driven insulating phases near commensuration. It must be noted that there are multiple counts on which our toy model is {\it not} a faithful description of TBLG. As such, it is not intended to be the last word on the description of the system. Nevertheless, we believe that it {\it does} capture the essential physics, particularly with regard to the effects of commensuration and Zeeman field, and as such may be a fruitful point of departure for future investigations. In particular, our analysis provides a natural way to incorporate the intuition of `superconductor as primary phase, interrupted by insulator at commensurate fillings,' which is suggested by the data of \cite{Lu2019Efetov}, and if this turns out to be the correct perspective on the problem, then the framework we have herein introduced may prove to be a natural point of departure for theoretical descriptions.  Finally, of course, the model we have introduced is interesting in its own right, exhibiting a rich phase diagram, and is therefore also itself worthy of experimental investigation.

\section*{Acknowledgments}
We thank Matthew Foster, Andrew Potter, Fengcheng Wu, and Rui-Xing Zhang for useful discussions. Y.-Z.C. thanks Leo Radzihovsky and Tzu-Chi Hsieh for collaborating on related projects in which some of the insights are invaluable. Research was sponsored by the Army Research Office and was accomplished under Grant Number W911NF-17-1-0482 (Y.-Z.C., Y.-P.L., and R.M.N.), by a Simons Investigator award from the Simons Foundation to Leo Radzihovsky (Y.-Z.C.), and the Laboratory for Physical Sciences and Microsoft (S.D.S.).
S.D.S. and R.M.N. would like to acknowledge the hospitality of the KITP, where this work was conceived during a visit to the program `Correlations in Moir\'e flat bands,' and also the Aspen Center for Physics, where the paper was completed during a visit to the program `Moiré Materials: Strong Correlations in Synthetic Superlattices.’ The KITP is supported in part by the National Science Foundation under Grant No. NSF PHY-1748958. The Aspen Center for Physics is supported by National Science Foundation grant PHY-1607611.
The views and conclusions contained in this document are those of the authors and should not be interpreted as representing the official policies, either expressed or implied, of the Army Research Office or the U.S. Government. The U.S. Government is authorized to reproduce and distribute reprints for Government purposes notwithstanding any copyright
notation herein.

\appendix

\section{Bosonization Convention}\label{App:Bosonization}

We apply the standard bosonization \cite{Shankar_Book,Giamarchi_Book} in this section. The right and left movers are bosonized via the following formulae.
\begin{align}
R_{\sigma}=\frac{\hat{U}_{\sigma}}{\sqrt{2\pi\alpha}}e^{i\left(\phi_{\sigma}+\theta_{\sigma}\right)},\,\,\, L_{\sigma}=\frac{\hat{U}_{\sigma}}{\sqrt{2\pi\alpha}}e^{i\left(\phi_{\sigma}-\theta_{\sigma}\right)},
\end{align}
where $\alpha$ is the ultraviolet length scale and $\hat{U}_{\sigma}$ is the Klein factor with spin $\sigma$. 
The density and the current of each spin sector can be expressed by $\rho_{\sigma}=\frac{1}{\pi}\partial_x\theta_{\sigma}$ and $I_{\sigma}=-\frac{1}{\pi}\partial_t\theta_{\sigma}$.
Following the standard treatment \cite{Giamarchi_Book}, we introduce the charge and spin collective variables as follows:
\begin{align}
\theta_{c}=&\frac{1}{\sqrt{2}}\left(\theta_{\uparrow}+\theta_{\downarrow}\right),\,\,\,\phi_{c}=\frac{1}{\sqrt{2}}\left(\phi_{\uparrow}+\phi_{\downarrow}\right)
,\\
\theta_{s}=&\frac{1}{\sqrt{2}}\left(\theta_{\uparrow}-\theta_{\downarrow}\right),\,\,\,\phi_{s}=\frac{1}{\sqrt{2}}\left(\phi_{\uparrow}-\phi_{\downarrow}\right).
\end{align}

The quantum wire with intrawire interactions, $H_U$ and $H_V$ [given by Eq.~(\ref{Eq:H_1D:int})], can be decomposed in the spin and charge collective coordinates.

\section{Bosonized impurity interactions}\label{App:RG_flow:imp}

The wire-wire impurity interactions [given by Eq.~(\ref{Eq:H_X:ww})] are bosonized to
\begin{subequations}\label{Eq:H_X:ww:bosonization}
\begin{align}
\nonumber H_{e}=&t_e\sum_{\sigma=\uparrow,\downarrow}\left[\left(R^{\dagger}_{\sigma}+L^{\dagger}_{\sigma}\right)_y\left(R_{\sigma}+L_{\sigma}\right)_x+\text{H.c.}\right],\\
\rightarrow&\frac{2t_e}{\pi\alpha}\!\sum_{\sigma=\uparrow,\downarrow}\!\!\hat{U}_{\sigma,y}\hat{U}_{\sigma,x}\left[e^{-i\phi_{\sigma}}\cos \theta_{\sigma}\right]_y\!
\left[e^{i\phi_{\sigma}}\cos \theta_{\sigma}\right]_x
\!\!+\!\!\text{H.c.}\\
\nonumber H_{C}=&t_u\sum_{\sigma,\sigma'}\!:\!\left(L^{\dagger}_{\sigma}R_{\sigma}+R^{\dagger}_{\sigma}L_{\sigma}\right)_y\!\left(L^{\dagger}_{\sigma'}R_{\sigma'}+R^{\dagger}_{\sigma'}L_{\sigma'}\right)_x:,\\
\rightarrow&\frac{4t_u}{\pi^2\alpha^2}\!\!\left[\cos\!\left(\!\sqrt{2}\theta_c\right)\!\cos\!\left(\!\sqrt{2}\theta_s\right)\right]_y\!\!\left[\cos\!\left(\!\sqrt{2}\theta_c\right)\!\cos\!\left(\!\sqrt{2}\theta_s\right)\right]_x\\
\nonumber H_{2e}=&t_{2e}\left[\left(R^{\dagger}_{\uparrow}L^{\dagger}_{\downarrow}-R_{\downarrow}^{\dagger}L^{\dagger}_{\uparrow}\right)_y\left(L_{\downarrow}R_{\uparrow}-L_{\uparrow}R_{\downarrow}\right)_x+\text{H.c.}\right]\\
\rightarrow&\frac{t_{2e}}{\pi^2\alpha^2}\hat{O}_{2e}
e^{i\sqrt{2}(\phi_{c,x}-\phi_{c,y})}\!\cos\!\left(\!\sqrt{2}\theta_{s,y}
\right)\!\cos\!\left(\!\sqrt{2}\theta_{s,x}
\right)\!+\!\text{H.c.}
\end{align}
\end{subequations}
where $\hat{O}_{2e}=\hat{U}_{\uparrow,y}^{\dagger}\hat{U}_{\downarrow,y}^{\dagger}\hat{U}_{\downarrow,x}\hat{U}_{\uparrow,x}$. In addition to the above perturbations, other impurity interactions can also give terms like $(\partial_x\theta_c)\cos(\sqrt{2}\theta_c)$. However, these perturbations are always irrelevant in the RG analysis. We also note that the $H_C$ interaction here is different from the crossed sliding Luttinger liquid study \cite{Mukhopadhyay2001}. Therefore, we don't expect sliding Luttinger phase here.

\section{Effective Josephson coupling}\label{App:EJC}

Here, we derive the effective Josephson coupling among two adjacent superconducting puddles. We follow the ideas listed in the Appendix of Ref.~\cite{Spivak2008}. 
We consider a 1D quantum wire with two superconducting puddles located at $x=0$ and $x=r$. The imaginary-time action is $\mathcal{S}_{\text{1D}}+\mathcal{S}_{\text{puddle}}+\mathcal{S}_I$ where $\mathcal{S}_{\text{1D}}$ describes a 1D spinful Luttinger liquid, $\mathcal{S}_{\text{puddle}}$ is the action of the two superconducting puddles, and
\begin{align}
\mathcal{S}_I=J\int\limits_{\tau}\left[\Delta^*_{0}(\tau)\mathcal{C}(\tau,0)+\Delta^*_{r}(\tau)\mathcal{C}(\tau,r)+\text{H.c.}\right]
\end{align}
is the Cooper pair hopping. $\Delta_x$ is the local superconducting complex order parameter at position $x$ and $\mathcal{C}$ is the singlet Cooper pair field. After integrating over the $\mathcal{S}_{\text{1D}}$, we can derive the lowest order [$O(J^2)$] effective Josephson coupling among puddles as follows
\begin{align}
\mathcal{S}_{J}=&-\frac{J^2}{2}\int\limits_{\tau,\tau'}\!\!\left[\Delta_{r}^*(\tau)\Delta_{0}(\tau')\left\langle\mathcal{C}(\tau,r)\mathcal{C}^{\dagger}(\tau',0)\right\rangle+\text{H.c.}
\right]\\
=&
-\frac{J^2}{2}\int\limits_{\omega_n}\left[\tilde{\Delta}_{r}^*(\omega_n)\tilde{\Delta}_{0}(\omega_n)\kappa(\omega_n,r;\beta)+\text{H.c.}\right],
\end{align}
where $\beta$ is the inverse temperature, $\tilde{\Delta}_x(\omega_n)$ is the Fourier transform of $\Delta_x(\tau)$, and
\begin{align}\label{Eq:K_omega_r}
\kappa(\omega_n,r;\beta)=\int_{0}^{\beta}d\tau\, e^{i\omega_n\tau}\left\langle\mathcal{C}(\tau,r)\,\mathcal{C}^{\dagger}(0,0)\right\rangle.
\end{align}
The long time dynamics is governed by $\kappa(\omega_n= 0,r;\beta)$. In addition, the effective Josephson coupling among two superconducting puddles is $J_{\text{eff}}=J^2|\Delta|^2\kappa(0,r;\beta)$.

\begin{figure}[]
	\includegraphics[width=0.375\textwidth]{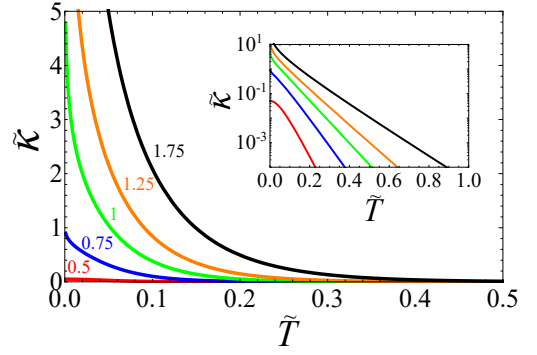}
	\caption{The effective Josephson coupling as a function of temperature. Based on Eq.~(\ref{Eq:kappa_0_T}), we define a dimensionless function $\tilde{\kappa}(\tilde{T},\tilde{r})\equiv \pi^2 v_c\alpha2^{1-\frac{1}{2K_c}}\kappa(0,r;\beta)$, where $\tilde{T}=\frac{\pi \alpha}{\beta v_c}$ and $\tilde{r}=r/\alpha$. We fix $\tilde{r}=20$ and plot $\tilde{\kappa}(\tilde{T},\tilde{r})$ for different values of the Luttinger parameter $K$. Red line indicates $K=0.5$; blue line indicates $K=0.75$; green line indicates $K=1$; orange line indicates $K=1.25$; black line indicates $K=1.75$. For all the cases, $\tilde{\kappa}$ is a monotonically decaying function in $\tilde{T}$. We note that the Luttinger liquid theory is valid for $\tilde{T}<1$ since $\tilde{T}$ is the ratio of physical temperature to the cutoff energy. Tuning $\tilde{r}$ will not change the qualitative results. Generally, larger $\tilde{r}$ means smaller $\tilde{\kappa}(\tilde{T},\tilde{r})$. Inset: The semi-log plot of $\tilde{\kappa}$ v.s. $\tilde{T}$.
	}
	\label{Fig:kappa_T}
\end{figure}

Within bosonization, the Cooper pair correlation function is given by
\begin{align}
\nonumber&\left\langle\mathcal{C}(\tau,r)\,\mathcal{C}^{\dagger}(0,0)\right\rangle\\
\nonumber=&\frac{1}{\pi^2\alpha^2} 
\left\langle e^{i\sqrt{2}\phi_c(\tau,r)} e^{-i\sqrt{2}\phi_c(0,0)}
\right\rangle_c\\
\label{Eq:Cooper_corr}&\times\left\langle \cos\left[\sqrt{2}\theta_s(\tau,r)
\right]\cos\left[\sqrt{2}\theta_s(0,0)
\right]
\right\rangle_s,
\end{align}
where the spin and charge sectors can be evaluated separately. For simplicity, we assume a spin gap and $V<1$. In the semiclassical approximation, the spin sector $\left\langle \cos\left[\sqrt{2}\theta_s(\tau,r)
\right]\cos\left[\sqrt{2}\theta_s(0,0)
\right]
\right\rangle_s\approx 1$. Therefore, the Cooper pair correlation function becomes \cite{Shankar_Book,Giamarchi_Book}
\begin{align}
\nonumber&\left\langle\mathcal{C}(\tau,r)\,\mathcal{C}^{\dagger}(0,0)\right\rangle\\
\approx &\frac{1}{\pi^2\alpha^2}
\left\{\frac{\left(\frac{\pi\alpha}{\beta v_c}\right)^2}{\sinh\left[\frac{\pi}{v_c\beta}(r+iv_c\tau)\right]\sinh\left[\frac{\pi}{v_c\beta}(r-iv_c\tau)\right]}\right\}^{\frac{1}{2K_c}}\\
=&\frac{1}{\pi^2\alpha^2}
\frac{2^{\frac{1}{2K_c}}\left(\frac{\pi\alpha}{\beta v_c}\right)^{\frac{1}{K_c}}}
{\left[\cosh\left(\frac{2\pi r}{v_c\beta}\right)-\cos\left(\frac{2\pi\tau}{\beta}\right)\right]^{\frac{1}{2K_c}}}.
\end{align}
With the above expression, the correlation function $\kappa(0,r;\beta)$ given by [Eq.~(\ref{Eq:K_omega_r})] can be computed using MATHEMATICA as follows:
\begin{align}
\nonumber&\kappa(0,r;\beta)\\
\nonumber=&\frac{2^{\frac{1}{2K_c}}\left(\frac{\pi\alpha}{\beta v_c}\right)^{\frac{1}{K_c}}}{\pi^2\alpha^2}\frac{\beta}{2\pi}\\
\nonumber&\!\!\!\times\!\left\{\!\frac{_2F_1\left(\frac{1}{2},\frac{1}{2K_c};1;\frac{-2}{A-1}\right)}{\left[A-1\right]^{\frac{1}{2K_c}}}
\!+\!\frac{_2F_1\left(\frac{1}{2},\frac{1}{2K_c};1;\frac{2}{A+1}\right)}{\left[A+1\right]^{\frac{1}{2K_c}}}\!
\!\right\}\\
\nonumber=&\frac{2^{\frac{1}{2K_c}-1}\tilde{T}^{\frac{1}{K_c}-1}}{\pi^2v_c\alpha}\\
\label{Eq:kappa_0_T}&\times\left\{\!\frac{_2F_1\left(\frac{1}{2},\frac{1}{2K_c};1;\frac{-2}{A-1}\right)}{\left[A-1\right]^{\frac{1}{2K_c}}}
\!+\!\frac{_2F_1\left(\frac{1}{2},\frac{1}{2K_c};1;\frac{2}{A+1}\right)}{\left[A+1\right]^{\frac{1}{2K_c}}}\!
\!\right\}
\end{align}
where $A=\cosh\left(2\tilde{T}\tilde{r}\right)$, $\tilde{T}=\frac{\pi \alpha}{\beta v_c}$ is the dimensionless temperature rescaled by the cutoff energy ($\sim v_c/\alpha$), $\tilde{r}=r/\alpha$ is the dimensionless length, and $_2F_1$ is the ordinary hypergeometric function. We further define a dimensionless function $\tilde{\kappa}(\tilde{T},\tilde{r})\equiv \pi^2 v_c\alpha2^{1-\frac{1}{2K_c}}\kappa(0,r;\beta)$.
In Fig.~\ref{Fig:kappa_T}, $\tilde{\kappa}(\tilde{T},\tilde{r})$ is a monotonically decaying function in temperature for the physically relevant parameter regime (e.g., $1/2\le K<2$). The critical temperature of the 2D superconductivity in the main text is determined by $T_c=J_{\text{eff}}(T_c)=J^2|\Delta|^2\kappa(0,r;1/T_c)$.

Another interesting question is how does the effective Josephson coupling varies in $r$ at low temperatures. We first take the zero temperature limit of Eq.~(\ref{Eq:Cooper_corr}) and then compute $\kappa(0,r;\infty)\propto\int_{0}^{\infty}d\tau(r^2+v_c^2\tau^2)^{-\frac{1}{2K_c}}$. For $K_c<1$, $\kappa(0,r;\infty)\propto |r|^{1-\frac{1}{K_c}}$, a power law decay function depending on $K_c$. Meanwhile, $\tilde{\kappa}(0,r;\infty)$ diverges for $K_c>1$. The divergence indicates a distance independent Josephson coupling in the effective action.
Therefore, the putative zero temperature superconducting order with $K_c>1$ is robust against finite range perturbations (including disorder).

Finally, we discuss the effective Josephson couplings but for spin-gapless quantum wires. $\kappa(0,r;\beta)$ for the general cases is difficult to compute analytically (except for $v_c=v_s$). However, we do expect similar monotonically decay behavior in temperature. The zero-temperature effective Josephson coupling can be deduced from the equal-time Cooper pair correlation, $\left\langle\mathcal{C}(0,r)\,\mathcal{C}^{\dagger}(0,0)\right\rangle\propto r^{-\frac{1}{K_c}-K_s}$. This implies that a divergent $\kappa(0,r;\infty)$ when $\frac{1}{K_c}+K_s<1$.


\begin{thebibliography}{50}
\bibitem[{fn1()}]{fn1}
	\bibinfo{note}{In the semiclassical limit with $V>0$, the pinning value
		$\theta_{s,0}=\pi n/\sqrt{2} +\pi/2\sqrt{2}$ leads to
		$\left\langle\cos\left(\sqrt{2}\theta_{s,0}\right)\right\rangle=0$. We note
		that the expectation value should not be actually zero due to the quantum
		mechanical fluctuation. ($\phi_s$ and $\theta_s$ do not commute.) On the
		other hand, $|\cos\left(\sqrt{2}\theta_s\right)|\approx 1$ for $V<0$.}
\bibitem[{fn3()}]{fn3}
	\bibinfo{note}{Note that SC$^*$ is the same phase as SC, the $^*$ simply
		denotes that it is forming in the region with repulsive intrawire
		interactions.}
\bibitem[{fn2()}]{fn2}
	\bibinfo{note}{Within the standard Luther-Emery refermionization analysis, the
		critical value of the Zeeman field such that the Zeeman energy equals to half
		of the spin gap.}
\end{thebibliography}



\end{document}